\documentclass{emulateapj}
\usepackage{natbib}
\newcommand{\msun}{\,\rm M_\odot}

\newcommand{\be}{\begin{equation}}
\newcommand{\ee}{\end{equation}}
\newcommand{\ba}{\begin{eqnarray}}
\newcommand{\ea}{\end{eqnarray}}

\newcommand{\vmax}{V_{\rm max}}

\newcommand{\kms}{\,{\rm km\;s}^{-1}}

\shorttitle{Infall caustics in dark matter halos?}
\shortauthors{Diemand \& Kuhlen}
\begin{document}
\title{Infall caustics in dark matter halos?}

\author{J\"urg Diemand\altaffilmark{1,2} and Michael Kuhlen\altaffilmark{3}}

\altaffiltext{1}{University of California, High Street 1156, Santa Cruz, CA 95064; diemand@ucolick.org}
\altaffiltext{2}{Hubble Fellow}
\altaffiltext{3}{Institute for Advanced Study, Einstein Drive, Princeton, NJ 0854; mqk@ias.edu}

\begin{abstract}
  We show that most particle and subhalo orbits in simulated
  cosmological cold dark matter halos are surprisingly regular and periodic: The
  phase space structure of the outer halo regions shares some of the
  properties of the classical self-similar secondary infall model.
  Some of the outer branches are clearly visible in the
  radial velocity - radius plane at
  certain epochs. However, they are severely broadened in
  realistic, triaxial halos with non-radial, clumpy, mass accretion.
  This prevents the formation of high density caustics:
  Even in the best cases there are only broad, very small ($<10\%$) enhancements
  in the spherical density profile. Larger fluctuations in $\rho(r)$ caused
  by massive satellites are common.
  Infall caustics are therefore too weak to affect lensing
  or dark matter annihilation experiments. 
  Their detection is extremely challenging,
  as it requires a large number of accurate tracer positions and
  radial velocities in the outer halo. The stellar halo of the Milky
  Way is probably the only target where this could become feasible in the future.
\end{abstract}


\section{Introduction}
Idealized models of self-similar, radial infall onto an initial
spherical overdensity in an expanding and otherwise homogenous
universe can be solved exactly
\citep{1984ApJ...281....1F,1985ApJS...58...39B} (FGB hereafter).  The
two key predictions of these models are: (i) Nearly isothermal
($\rho \propto r^{-2}$) halo density profiles, as found in earlier models
\citep{1975ApJ...201..296G,1977ApJ...218..592G}. (ii) Near their
apocenters mass shells pile up and form spherical caustic surfaces of
infinite density. This occurs at constant fractions of the {\it
current} turnaround radius, resulting in slowly outward moving
caustics as the turnaround radius increases with time. Each mass shell
moves on a radial orbit with a nearly fixed (only slowly decreasing)
apocenter of about 0.85 of {\it its own} turnaround radius. After a
shell falls in it forms part of the first (outermost) caustic 
when the shell is near its first apocenter passage.
Later, near its second apocenter, it forms part of the second caustic,
and so on.  Such caustics would increase the dark matter annihilation
rate
\citep{2001PhRvD..63h3515B,2006MNRAS.366.1217M}, and it has been
argued that infall would lead to peaks in the local CDM
velocity distribution and change the predictions for direct dark
matter detection \citep{1997PhRvD..56.1863S}.

\begin{figure*}
\plotone{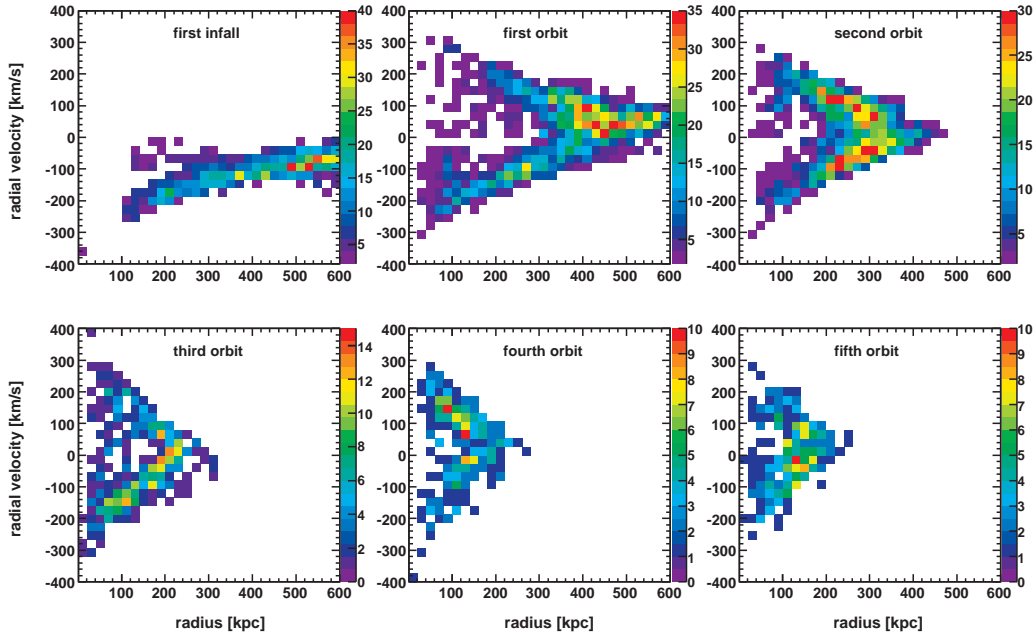}
\caption{Phase space distribution of subhalos at z=0 separated by the number of orbits
they have completed. This figure is also available as an mpeg
animation running from z=12 to 0 in the electronic edition of ApJ.}
\label{fig:subhalo_caustics}
\end{figure*}

Both key predictions of the original FGB-model are at odds with the
properties of dark matter halos found in cosmological simulations:
Halo density profiles are found to be steeper than $\rho \propto
r^{-2}$ in the outer parts and shallower in the inner regions
\citep{1991ApJ...378..496D,1996ApJ...462..563N}.  However, by relaxing
some of the simplifying assumptions of the FGB-model it is possible to
obtain a more realistic secondary infall model which successfully
reproduces the halo density profiles found in cosmological simulations
(e.g. \citealt {2007MNRAS.376..393A} and references therein):
allowing for angular momentum lowers the inner densities relative to the original
$\rho \propto r^{-2.25}$ profile,
simply because fewer particles orbit through the central regions \citep{1993ApJ...418....4R}.
More realistic initial perturbations lead to a steeper outer profile \citep{1985ApJ...297...16H}.

Infall does not lead to high density caustic surfaces in cosmological simulations.
In early simulations most particle orbits
differed widely from the slowly shrinking, nearly
periodic orbits found in the FGB-model \citep{1996ApJ...457...50Z}.
But these simulations contained only a few thousand particles,
which leads to significantly distorted orbits due to
artificial two body relaxation \citep{2004MNRAS.348..977D}.
Despite the numerical noise, \citet{1996ApJ...457...50Z} found
that the final particle energies are strongly correlated with
the initial ones, suggesting a rather gentle, orderly halo build-up as in FGB,
and not violent relaxation. Another consequence of this fairly regular build-up is
that material from biased, early high density regions
ends up in the inner parts of halos today \citep{2005MNRAS.364..367D}.
However, recent high resolution simulations 
still produce rather smooth mass profiles without high density caustics 
(e.g. \citealt{2001PhRvD..64f3508M,2002PhRvD..66f3502H,2007ApJ...667..859D}).
\citet{2001idm..conf...93M} emphasized that in the hierarchical
buildup of CDM halos the internal velocity dispersion of
infalling satellites $\sigma_{\rm sat}$
is typically only a few times smaller than $\sigma_{\rm host}$,
which would broaden caustics significantly, to a width of about
$R_{\rm host}\sigma_{\rm sat} \, / \, \sigma_{\rm host}$. Indeed
resolved mergers do bring in practically all the mass and
there is no evidence for smooth accretion
\citep{2008arXiv0802.2265M}.  Here we show that
despite clumpy, non-radial, anisotropic accretion into a non-spherical potential,
the outer parts of halos do show a few weak, broad branches
in $v_r - r$-space. Their properties agree well with some of the FGB
predictions. In contrast to the model however, realistic infall
"caustics"\footnote{For brevity we refer to the turnaround regions
of branches in $v_r - r$-space simply as "caustics", even tough they
don't have the high densities implied by the true sense of the word.}
increase the dark matter densities only slightly ($\simeq
10\%$ for the strongest ones). Unfortunately this makes them
practically undetectable and irrelevant for dark matter detection.

\section{Weak outer caustics in $v_r - r$ space}

\begin{figure*}
\plotone{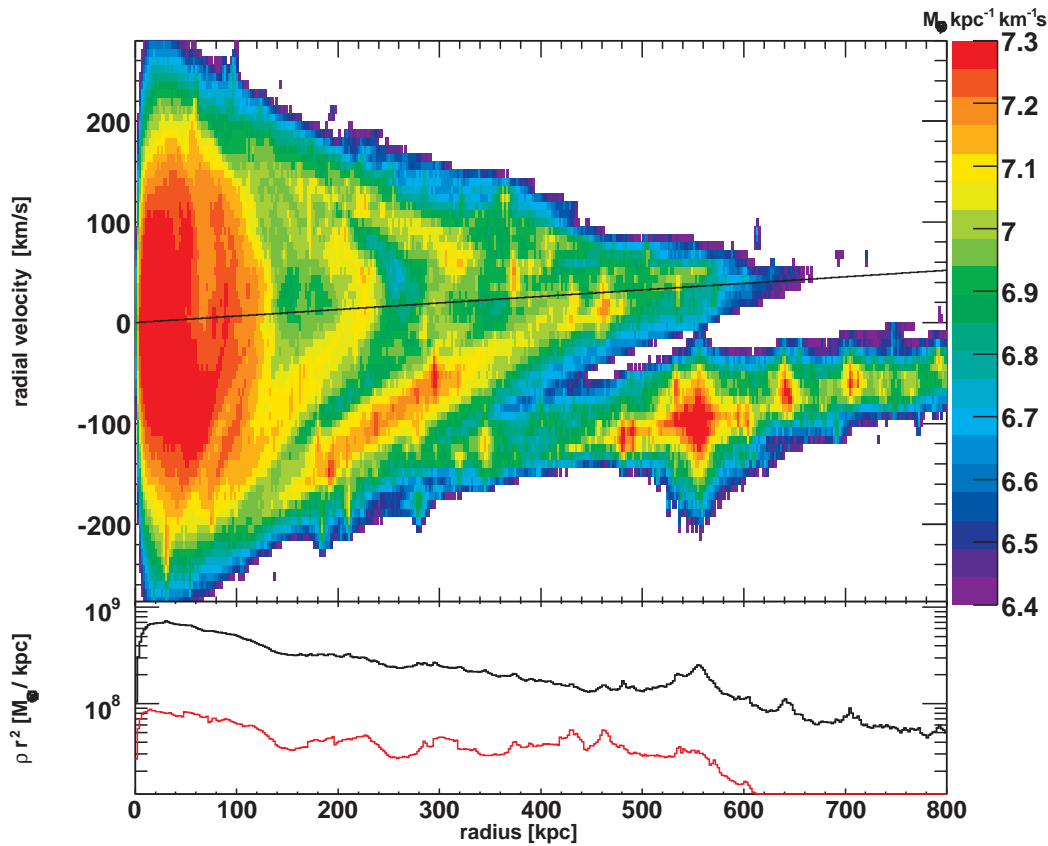}
\caption{Phase space distribution of the dark matter particles at z=0.  The straight
  line is $v_{\rm caustic}(r)$. The lower panel shows the $\rho(r) \,
  r^2$ of all particles (black) and of a subset with $|v_r-v_{\rm
    caustic}| < 24 \kms$ (red).  This figure is also available as an
  mpeg animation running from z=12 to 0 in the electronic edition of
  ApJ.}
\label{fig:particle_caustics}
\end{figure*}

In this section we compare the histories of subhalos in the Via Lactea simulation
(\citealt{2007ApJ...667..859D}, VL07 hereafter)\footnote{Subhalo
histories, animated versions of Figures~\ref{fig:subhalo_caustics}
and \ref{fig:particle_caustics} and other movies are available at
www.ucolick.org/$\sim$diemand/vl} with predictions from the
FGB-model. Later we will include the distribution of all dark matter
particles. The subhalos are selected by peak circular velocity $\vmax
> 3 \kms$ {\it before} accretion, and are therefore very similar to
dark matter particles in their spatial and velocity distribution
\citep{2006MNRAS.369.1698F}. This sample contains 10,652 
halos and subhalos with masses above about $3 \times 10^6 \msun$. The host
halo has size of $\vmax= 181 \kms$, consistent with the dark halo of the Milky Way.
The subhalos which have completed the
same small number of orbits end up in a relatively narrow branch in
$v_r - r$ space, despite their range of orbital eccentricities (VL07),
the prolate host halo potential (about
 0.8:1, see \citealt{2007ApJ...671.1135K}) and occasional kicks from larger
satellites \citep{2007MNRAS.379.1475S}.  Each panel in
Figure~\ref{fig:subhalo_caustics} would correspond to one
infinitesimally thin branch in the $v_r - r$ plane of the FGB-model.
Already at first infall the radial and tangential velocity dispersions
are quite large: $\sigma_r \simeq 26 \kms$ and $\sigma_{\rm tan,2D}
\simeq 60 \kms$.  These motions induced by large scale tidal
forces and also by structures forming within the turnaround region
prevent the formation of thin, high density caustics.

\begin{deluxetable*}{l|ccc|ccc|ccc|cc} 
\tabletypesize{\scriptsize}
\tablecaption{Positions and widths of caustics and turnaround regions in the Via Lactea halo.}
\tablewidth{0pt}
\tablehead{
\colhead{k}&\colhead{$r_{k,{\rm med}}$}&\colhead{$r_{k,68\%}$}&
\colhead{$\frac{\Delta r_{k}}{r_{k,{\rm med}}}$}&
\colhead{$t_{k,{\rm med}}$} & \colhead{$t_{k,68\%}$}&
\colhead{$\frac{\Delta t_{k}}{t_{k,{\rm med}}}$}&
 \colhead{$\left(\frac{r_{k}}{t_{k}}\right)_{\rm med}$}&
 \colhead{$\left(\frac{r_{k}}{t_{k}}\right)_{68\%}$}&\colhead{$\left(\frac{r_{k}}{t_{k}}\right)_{\rm FGB}$} 
 &
 \colhead{$\frac{r_{k,{\rm med}}}{r_{1,{\rm med}}}$} & \colhead{$\left(\frac{r_{k}}{r_{1}}\right)_{\rm FGB}$
}  \\
\colhead{ } & \colhead{[kpc]} & \colhead{[kpc]} & \colhead{ } & \colhead{[kpc]} & \colhead{[kpc]} 
& \colhead{ }& \colhead{ }& \colhead{ } }
\startdata
1&453&370$-$534&0.36&491&443$-$551&0.22&0.92&0.77$-$1.12&0.876&1&1\\
2&310&242$-$384&0.46&343&297$-$407&0.32&0.93&0.57$-$1.24&0.864&0.68&0.65\\
3&220&204$-$237&0.15&261&211$-$316&0.40&0.84&0.67$-$1.10&0.856&0.49&0.49\\
4&173&137$-$207&0.41&222&180$-$266&0.39&0.78&0.58$-$1.25&0.843&0.38&0.40\\
5&141&110$-$191&0.57&179&131$-$229&0.55&0.78&0.52$-$1.46&0.832&0.31&0.34\\
6&121& 89$-$170&0.67&157&105$-$201&0.61&0.81&0.54$-$1.46&0.834&0.27&0.30\\
\enddata

\tablecomments{Positions and widths of the outer six caustics $r_{k}$
  and turnaround radii of the particles in the caustic $t_{k}$. Median values and $68\%$ ranges
  of all caustic members are given, $\Delta$ refers to the full width of the $68\%$ range.  The number
  of subhalo members in the caustics 1 to 6 are 551, 300, 49, 32, 56 and 15. "FGB"
  refers to predictions from the secondary infall model.}
\label{table:caustics}
\end{deluxetable*}

From the positions of subhalos near their apocenter we can measure the
position and width of the outer six infall caustics
(Table~\ref{table:caustics}). In the FGB-model the radial velocity of
a caustic at radius $r$ is $v_{\rm caustic}(r) = 8/9 \; r/t$, where
$t$ is the time since the Big Bang\footnote{For simplicity we assume
this formula derived by FGB in an $\Omega_m = 1$ universe to be
approximately right also in $\Lambda$CDM.}.  We define those
subhalos near their k$^{\rm th}$ apocenter passage and with a radial
velocity within $24 \kms$ of $v_{\rm caustic}(r)$ as the members of
caustic k. For each member $i$ we then determine its turnaround
radius $t_{k,i}$ and the ratio of today's position to
turnaround radius $r_{k,i}/t_{k,i}$. Note that members of one caustic
fell in from all directions, not as a bound group, and merely share
similar turnaround times and radii. The scatter in the turnaround
radii is relatively large and comparable to the width of the caustics.
This again hints at a broadening of caustics caused already early
(near the turnaround time) by proper motions. Caustics 4, 5, and 6 are
so wide that they completely overlap, and they disappear in the full
$v_r - r$ plot of all particles (Figure~\ref{fig:particle_caustics}).
Caustic 3 is the most prominent one at z=0 and it leads to a small
over-density of about 10\% over a smooth density profile around 220
kpc. The abundance of particles with $|v_r-v_{\rm caustic}| < 24 \kms$
increases by a factor of about 1.6 around 220 kpc. The outermost
caustics 1 and 2 are quite wide, but parts of their corresponding branches
from Figure 1 are still visible in Figure~\ref{fig:particle_caustics}. 
Note that all of the outer three $v_r - r$ branches have large gaps, which are caused by epochs of little or
no mass accretion. The animated versions of
Figures~\ref{fig:subhalo_caustics} and ~\ref{fig:particle_caustics}
shows how these phase-space features and the gaps in between form and
move. Whether a certain branch is currently visible in the $v_r - r$
plane depends on whether there was enough mass falling in at some
earlier time to reach a sufficient density in the corresponding region
of phase-space. The clearest caustic at z=0 is the third one; it was
generated by continuous, large infall of material around z=1.7. At
later epochs Via Lactea's mass accretion becomes quite small and
sporadic (VL07), leading to weak first and second branches with large
gaps.  The Via Lactea halo formed in a series of major mergers before
z=1.7. These mergers could be responsible for a broadening of the
fourth and higher caustics to a degree where they become completely
washed out and undetectable in the z=0 snapshot.

\section{Comparison with the FGB-model} 

In Table~\ref{table:caustics} we make some quantitative comparisons
with the $\epsilon=1$ version of the FGB-model, which corresponds to
a point-mass as initial over-density: $\delta = \delta M / M \propto M^{-\epsilon}$.
In CDM the initial peaks are extended. The Via Lactea halo has
$\epsilon \simeq 0.2$ up to about $10^{11} \msun$. For larger masses $\epsilon$
increases and it is about one at $10^{12} \msun$, equivalent
to a scale of 150 kpc today. The outer halo features discussed here
correspond to the $\epsilon\simeq1$ regime\footnote{Larger collapse factors
in the inner halo (see Fig. 1 in VL07) might be related to the smaller $r_{k}/t_{k}$ ratios
in the $\epsilon=0.2$ FGB-model.}.

The FGB-model was solved for an Einstein-de Sitter
($\Omega_m = 1$) universe, while the Via Lactea simulation assumed the
now standard $\Lambda$CDM model, but we do not believe that this
difference affects our results: In the model, the position of a
caustic is similar (0.8 to 0.9) to the turnaround radius of the
caustic members. Members of the first caustic turned around before z=1
(earlier for higher caustics), when the $\Lambda$CDM model was still
matter dominated and similar to an Einstein-de Sitter universe. After
turnaround the enclosed mass dominates over the enclosed vacuum energy
and the dynamics are practically the same in both cosmologies. 

We find that the ratios of the current caustic positions are in good
agreement with the FGB-model (Table~\ref{table:caustics}).  Also the
median of the ratios of caustic radii to turnaround radii $(r_{k} /
t_{k})_{\rm med}$ agree quite well with the FGB-model, i.e. the
typical orbits go out close to their turnaround scale. Earlier models
\citep{1977ApJ...218..592G} assumed they would decrease by a factor of
two during virialization. This assumption is also made in the
definition of the formal virial radius\footnote{The virial radius of the Via Lactea halo is
$r_{\rm vir}=r_{104}=288$ kpc. It encloses a mass of
$M_{\rm vir}=1.5 \times 10^{12} \msun$.}, which therefore only
encloses a fraction of the material orbiting around a galaxy
(\citealt{2006ApJ...645.1001P},VL07).  The $(r_{k} / t_{k})_{\rm med}$
in the clusters from \citep{2004MNRAS.352..535D} are also close to
0.85, i.e the orbits of bound material extend well beyond the formal
virial radius both in galaxy and cluster halos\footnote{Unlike
  galaxies halos, typical cluster halos do accrete a lot of mass
  around $z=0$ leading to a net infall between their formal virial
  radius and the turnaround radius \citep{2007arXiv0710.5520C}.}.

The main difference between the FGB-model and cosmological halos is
the large scatter in space and velocity around the lines in $v_r - r$
space on which all matter is found in the model. The scatter prevents
the formation of high density caustic surfaces and makes the broad
caustic-like features from cosmological infall practically
undetectable: To find them one needs an accurate, large sample of $v_r
- r$ measurements, which are available in simulations, but
unfortunately not in the real Universe.  For members of the outer four
caustics the scatter in the individual $r_{k} / t_{k}$ ratios is
consistent with the shape of the potential in the outer parts of Via
Lactea halo, which is about 1:0.8 prolate at z=0 and about 1:0.7
prolate at z=0.5 \citep{2007ApJ...671.1135K}.  We also find a few
outliers with much larger $r_{k} / t_{k}$ ratios, probably resulting
from dynamical interactions between multiple satellites
\citep{2007MNRAS.379.1475S}. The scatter is
significantly larger for the 5th and 6th caustics, presumably because
this material was accreted early ($z>1.7$), when a series of major
mergers caused large fluctuations in the host potential.

\section{Multimodal radial velocity distributions}

\begin{figure}
\epsscale{1.15} 
\plotone{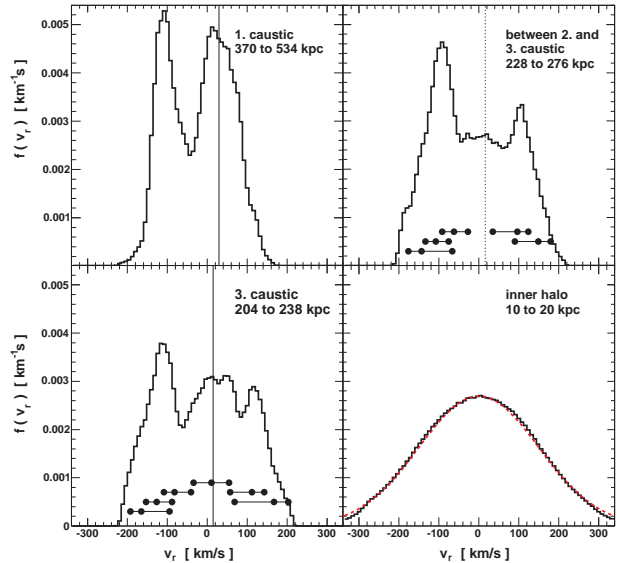}
\caption{Radial velocity distributions of particles in different radial shells (solid lines). Only in the inner halo
(lower right panel) the distribution is uni-modal and close to a Gaussian (dashed line). The other
panel show multi-modal distributions. To illustrate how to decompose these particle distributions 
we also plot the median and 68\% ranges of ingoing and outgoing subhalos,
which have completed 0, 1 or 2 two orbits from bottom to top (circles). On the third caustic (lower left panel)
the subhalos which have completed 3 orbits are now near their apocenter. The vertical lines give the local caustic
velocity $v_{\rm caustic}(r) = 8/9 \; r/t$.}
\label{fig:vr_distribution}
\end{figure}

From Figure~\ref{fig:particle_caustics} it is clear that the radial
velocity distribution of particles in a shell in the outer halo will
not be smooth and unimodal, but rather a superposition of modes caused
by the various branches in $v_r - r$ space.
The velocity distribution and relative importance of each mode changes with distance from the
galactic center, and fewer modes are present at large radii
(Figure ~\ref{fig:vr_distribution}).

The subhalo radial velocity distributions are similar to those of
particles and we can use subhalo histories to illustrate how the
particle distributions divide up into their components.  Between
caustics 2 and 3 one finds a broad, doubly peaked distribution. The
negative velocity peak is a superposition of first infall and subhalos
approaching the center on the second half of their first or second
orbit. The subhalos on first infall have the most negative typical
velocities, followed by those on their first and second orbit, in
qualitative agreement with the FGB-model. In contrast to the model,
we find these components have wide radial velocity distributions which
largely overlap and give rise to only one broad infalling peak. On the
third caustic the broad, double peaked distribution from in- and
outgoing material is the same as above, but now
there is an additional peak at low radial velocity from particles near
their third apocenter.

In the inner halo particles which completed
the same number of orbits are distributed so widely that the total
radial velocity distribution becomes one
featureless peak. It is close to a Gaussian distribution with zero
mean velocity and a dispersion of $150 \kms$. But note that a smooth, inner
$f(r,v_{r})$ does not preclude the existence of coherent structures in
the full phase-space density $f(\vec{x},\vec{v})$, such as bound
clumps (subhalos) and their unbound debris (tidal streams)
\citep{2001PhRvD..64f3508M,2002PhRvD..66f3502H}.

\section{Summary}

We have discovered broad, caustic-like features in $f(r,v_{r})$ in the
outer parts of CDM halos from cosmological simulations. They are too
broad to cause significant density enhancements and would therefore be
very challenging to detect. Since a large number of accurate distances
radial velocities are needed, we believe that the stellar halo of the
Milky Way is the only system where one might possibly detect such
features in the foreseeable future.

The basic properties of infall caustics are independent of redshift and
halo mass; they are similar in the Via Lactea galaxy halo and in
the cluster halos from \citet{2004MNRAS.352..535D}.  The galaxy halos
in that work were resolved with only a few million particles per halo
and clearly show similar features. A new, one billion particle galaxy
halo (Diemand et al. in prep.) also only shows broad, weak outer caustics.
Since the features described here do not change qualitatively when the
numerical resolution is increased by almost three orders of magnitude
we conclude that they are well resolved in simulations with a few
million particles per halo.

\acknowledgments

We thank Ed Bertschinger and Roya Mohayaee for helpful discussions 
and Peter Goldreich, Roya Mohayaee and Scott Tremaine for helpful
comments on an earlier version of this letter. We acknowledge support from NASA through a
Hubble Fellowship grant (JD) and and from the Hansmann
Fellowship at the Institute for Advanced Study (MK).

\bibliographystyle{apj}
\bibliography{../References/db} 
\end{document}